\documentclass[prl]{revtex4}

\usepackage[dvips]{graphicx}
\begin{document}

\title{Block-Layer Concept for the Layered Co Oxide: \\
A Design for Thermoelectric Oxides}

\author{Takenori Fujii}
\email{fujii@htsc.phys.waseda.ac.jp}
\author{Ichiro Terasaki}
\affiliation{Department of Applied Physics, School of Science and Engineering, \\
Waseda University, Tokyo 169-8555, Japan}

\date{\today}

\maketitle

\section{Introduction}

A thermoelectric material is a material which generates electricity from heat through the Seebeck effect, and pumps heat through the Peltier effect. The thermoelectric conversion efficiency is characterized by the figure of merit $Z=S^2/\rho\kappa$. ($S$, $\rho$, and $\kappa$ are thermopower, resistivity, and thermal conductivity.) In order to attain the high thermoelectric efficiency, thermoelectric materials need large thermopower, low resistivity and low thermal conductivity. In conventional thermoelectric materials, however, $S$, $\rho$, and $\kappa$ cannot be controlled independently, because they depend on the carrier density. The thermoelectric performance is optimized near a carrier density of 10$^{19}$ cm$^{-3}$, which is a typical value for a degenerate semiconductor. Actually, the thermoelectric materials were mainly searched in the degenerate semiconductors of high mobility. 

According to the conventional thermoelectrics, oxides have been considered to be unsuitable for thermoelectric materials, because they have a low mobility. However, the layered Co oxide NaCo$_2$O$_4$ was found to show large thermopower with low resistivity (100 $\mu$V/K and 200 $\mu$$\Omega$cm at room temperature, respectively)~\cite{terra1}. After the discovery of NaCo$_2$O$_4$, a lot of efforts have been devoted to exploring new Co oxides of high thermoelectric performance, and some layered Co oxides, such as Bi$_2$Sr$_2$Co$_2$O$_y$~\cite{BiCo,ito,raveau}, Ca$_3$Co$_4$O$_9$~\cite{CaCo1,CaCo2,masset,miya2,sebastien}, and TlSr$_2$Co$_2$O$_y$~\cite{TlCo}, are found to show good thermoelectric properties as well. We think that the large thermopower in the layered Co oxide cannot be explained by a conventional band model, and that the strong electron-electron correlation plays an important role in the large thermoelectric power~\cite{terra2}.

These layered Co oxides have a CoO$_2$ layer, which consists of edge-shared CoO$_6$ octahedra, as a common structural unit responsible for the electric conduction and the large thermopower. This is very similar to the crystal structure of high-$T_c$ superconducting Cu oxides, which have a CuO$_2$ layer as a common unit responsible for the metallic conduction and the high-$T_c$ superconductivity. In the material design for high-$T_c$ Cu oxides, the ``block-layer'' concept is well established~\cite{tokura}. Because of the structural similarity, we apply the block-layer concept to the layered Co oxide, and propose that the thermoelectric properties can be finely tuned by modifying the block layer.

The purpose of this paper is to introduce a new guideline to thermoelectric material design using the block-layer concept. Here we choose the Bi-based Co oxides Bi$_{2-x}$Pb$_x$$M$$_2$Co$_2$O$_y$ ($M$ = Sr, Ba) as a prime example to show how the block-layer concept works well. The Bi-based oxides have a wide family of layered compounds, which have the same block layers as the superconducting Cu oxide Bi$_2$Sr$_2$CaCu$_2$O$_{8+\delta}$. We further note that they have a great advantage that a large single crystal can be easily grown by traveling solvent floating zone (TSFZ) method.

The contents of this paper are as follows: In $\S$ 2, we introduce the concept of the block layer in high-$T_c$ Cu oxides, and explain their fundamental role of the block layer using some examples. In $\S$ 3, we apply the block layer concept to the thermoelectric oxides, and show differences between high-$T_c$ Cu oxides and the layered Co oxides. In $\S$ 4, we discuss the crystal structure and the substitution effects of the Bi-based Co oxides. In $\S$ 5, we discuss the effect of the block layer on the thermoelectric properties. Finally, we summarize the present paper in $\S$ 6.

\section{Block-Layer Concept for High-$T_c$ Copper Oxides}

Let us begin with the block-layer concept proposed for high-$T_c$ Cu oxides~\cite{tokura}. High-$T_c$ Cu oxides consist of the alternating stack of the superconductive CuO$_2$ layer and the other components called the ``block layer''. As a result, each CuO$_2$ layer is separated by the block layer to form a two-dimensional conductive sheet. Various combinations of the block layers and the CuO$_2$ layer make a large family of high-$T_c$ Cu oxides. The most important feature is that any combination of block layers can induce high-$T_c$ superconductivity in the CuO$_2$ layer, while crystal symmetry, chemical properties, and mechanical properties are completely different for different block layers. Thus, we can install an additional function to the high-$T_c$ Cu oxides by combining the block layer(s) without losing high-$T_c$ superconductivity in the CuO$_2$ layer. In this sense, the block-layer concept is similar to the functional-group concept in organic materials.

\begin{figure}
\centering{
\includegraphics[width=12cm]{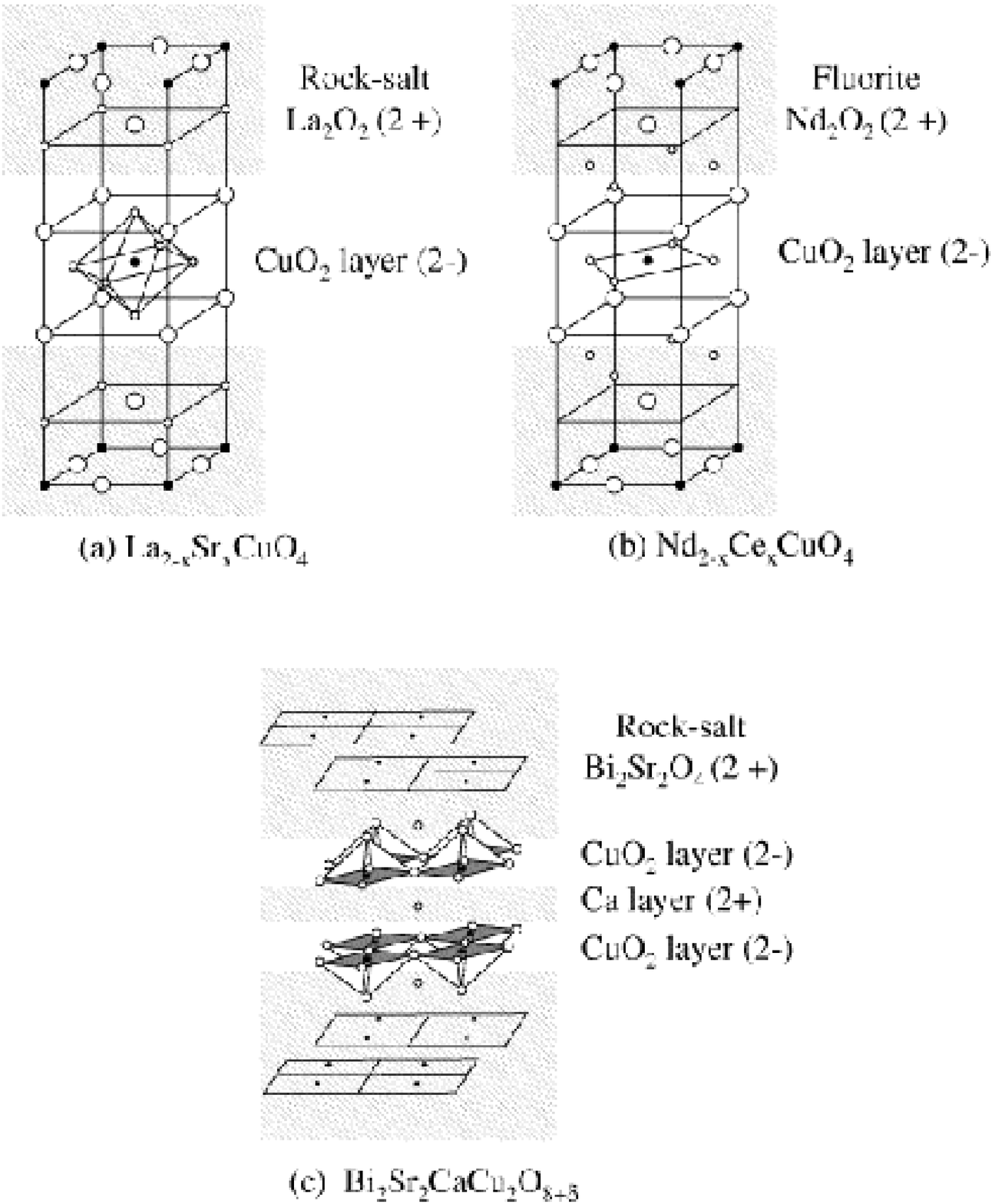}
\vspace{1cm}
}
\caption{
Crystal structure of (a) La$_{2-x}$Sr$_x$CuO$_4$, (b) Nd$_{2-x}$Ce$_x$CuO$_4$, and (c) Bi$_2$Sr$_2$CaCu$_2$O$_{8+\delta}$.
 }
\label{fig1}
\end{figure}

The stack of the CuO$_2$ layer and the block layer is schematically shown in Fig.~\ref{fig1}, where a rock-salt block, a fluorite block and a cation block are shown as examples. La$_{2-x}$Sr$_x$CuO$_4$ [Fig.~\ref{fig1}(a)] and Nd$_{2-x}$Ce$_x$CuO$_4$ [Fig.~\ref{fig1}(b)] have one kind of block layer, which corresponds to the rock-salt La$_2$O$_2$ layer and the fluorite Nd$_2$O$_2$ layer, respectively. In Bi$_2$Sr$_2$CaCu$_2$O$_{8+\delta}$ [Fig.~\ref{fig1}(c)], there are two kinds of block layers in a unit cell. One block layer is the rock-salt Bi$_2$Sr$_2$O$_4$ layer, and the other layer is the Ca layer. Then Bi$_2$Sr$_2$CaCu$_2$O$_{8+\delta}$ can be represented as doubly periodical sequence of -[Bi$_2$Sr$_2$O$_4$]-[CuO$_2$]-[Ca]-[CuO$_2$]-.

An important role of the block layer is to supply charge carrier in the CuO$_2$ layer. According to a simple valence counting, an insulating CuO$_2$ layer consists of Cu$^{2+}$ and O$^{2-}$, and hence is charged 2- ([CuO$_2$]$^{2-}$). To maintain the charge neutrality in whole crystal, the average charge of the block layer in the insulating (non-doped) compound must be 2+. Actually, the charge of the block layers La$_2$O$_2$ and Nd$_2$O$_2$ is 2+. The partial substitution of divalent Sr [tetravalent Ce] for trivalent La [Nd] in La$_{2-x}$Sr$_x$CuO$_4$ [Nd$_{2-x}$Ce$_x$CuO$_4$] introduces the excess charges of +$x$ [-$x$] into the CuO$_2$ layer. In the case of Bi$_2$Sr$_2$CaCu$_2$O$_{8+\delta}$, the charge of the block layers Bi$_2$Sr$_2$O$_4$ and Ca is 2+, and excess oxygen $\delta$ in the Bi$_2$O$_2$ plane supplies holes in the CuO$_2$ layer. Thus, the total valence in the block layers controls the carrier concentration in the CuO$_2$ layer. Note that we can dope the p-type and n-type carriers by choosing a proper block layer.

Another feature of the block layer is that the number of the CuO$_2$ layers in the unit cell can be changed with changing the sequence of the block layer. For example, the Bi-based Cu oxide Bi$_2$Sr$_2$Ca$_{n-1}$Cu$_n$O$_{2n+4}$ is composed of the stacking of the rock-salt Bi$_2$Sr$_2$O$_4$ and the CuO$_2$ layers separated by the Ca layer. Empirically, a high-$T_c$ Cu oxide with three CuO$_2$ layers in a primitive cell shows the highest $T_c$ in the homologous series~\cite{amaeda}.

\section{Block-Layer Concept for The Thermoelectric Cobalt Oxides}

\begin{figure}
\centering{
\includegraphics[width=12cm]{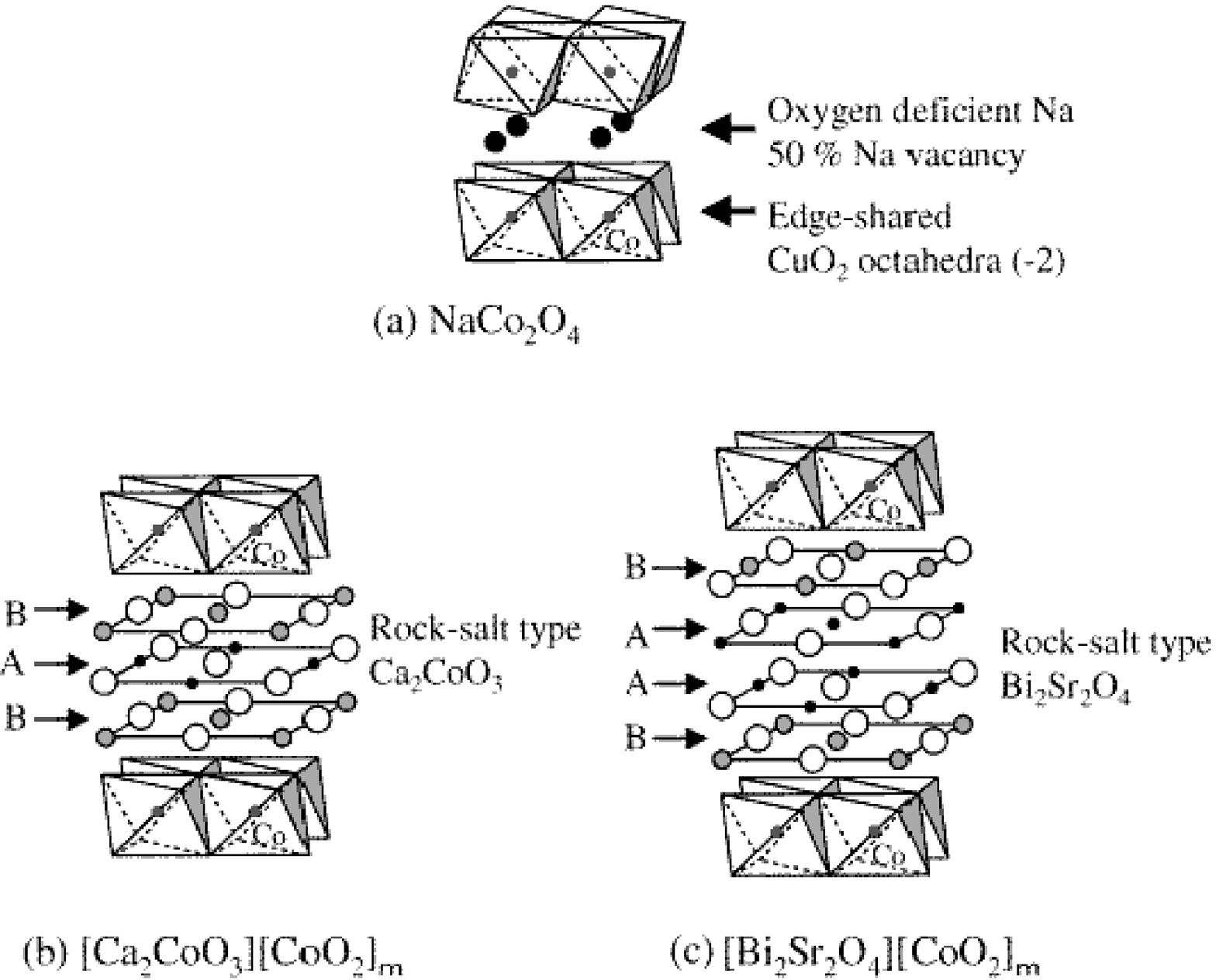}
\vspace{1cm}
}
\caption{
Crystal structure of (a) NaCo$_2$O$_4$, (b) [Ca$_2$CoO$_3$][CoO$_2$]$_m$, and (c) [Bi$_2$Sr$_2$O$_4$][CoO$_2$]$_m$.
 }
\label{fig2}
\end{figure}

Now, let us apply the block-layer concept to the layered Co oxide. The crystal structures of the layered Co oxides are shown in Fig.~\ref{fig2}. The simplest block layer is a cation lattice in NaCo$_2$O$_4$ (Na$_{0.5}$CoO$_2$) represented as -[Na$_{0.5}$]-[CoO$_2$]-, which is similar to the Ca layer in the Bi-based Cu oxide. Note that the Na layer is a highly disordered triangular lattice, in which about 50 \% Na sites are vacant. The thermal conductivity of the polycrystalline sample is as quite low as 15 mW/cmK$^2$~\cite{takahata1,takahata2}, which is due to the Na vacancy.

Next, we discuss the layered Co oxides with lattice misfit, which consist of an alternating stack of the rock-salt type block layer and the hexagonal CoO$_2$ conducting layer. These two layers have common the $a$ and $c$ axes, while the $b$-axis lengths of the two layers are different. ($b_{RS} \ne b_H$, where the subscripts RS and H refer to the rock-salt and hexagonal layers, respectively.) The block layer is expressed as [MO]$^{RS}_l$, where M = Pb, Hg, Tl, Bi, Ca, Co, and Sr, and $l$ is the number of MO layers in the rock-salt layer. The general formula of the misfit Co oxides can be expressed as [MO]$^{RS}_l$[CoO$_2$]$_m$, where $m$ = $b_{RS}/b_H$. The Bi-based Co oxide Bi$_2$Sr$_2$Co$_2$O$_y$~\cite{raveau} contains the quadruple rock-salt layer ($l$ = 4) [Fig.~\ref{fig2}(c)], and is expressed as [BiSrO$_2$]$^{RS}_2$[CoO$_2$]$_m$. The Ca-based Co oxide Ca$_3$Co$_4$O$_9$ contains the triple rock-salt layer ($l$ = 3) [Fig.~\ref{fig2}(b)], and is expressed as [Ca$_2$CoO$_3$]$^{RS}$[CoO$_2$]$_m$. The chemical compositions taken from the literature are listed in the Table~\ref{tbl1}, where A corresponds to the cation in the inner rock-salt layers, and B to that in the outer rock-salt layers. All the compounds are well described as [MO]$^{RS}_l$[CoO$_2$]$_m$. The Tl-based Co oxide~\cite{TlCo} has the triple rock-salt layer ($l$=3) with the formula [TlSr$_2$O$_3$]$^{RS}$[CoO$_2$]$_m$. Recently, the Pb-based Co oxide [Pb$_{0.7}$A$_{0.4}$Sr$_{1.90}$O$_3$]$^{RS}$[CoO$_2$]$_{1.8}$ (A = Hg, Co)~\cite{pelloquin} is discovered. The A site of the Tl- and Pb-based Co oxides is somewhat complicated, where a small amount of Co ions are mixed in the block layers.  Note that $b_{RS}$ of a Ca-based Co oxide is slightly smaller than the other ones ($b_{RS} \sim$ 4.5 \AA), suggesting a larger misfit. 

Although the block layers of these misfit layered Co oxides are very different in chemical properties, they exhibit very similar physical properties. The resistivities along the CoO$_2$ layer are 5 $\sim$ 30 m$\Omega$cm at room temperature~\cite{BiCo,ito,yamamoto,fujii1}, and show an upturn in the low temperature below about 50 K except for the Tl-based Co oxides~\cite{TlCo}. The thermopowers are as large as 80 $\sim$ 160 $\mu$V/K at room temperature. Their thermal conductivities are very low (typically 10 $\sim$ 20 mW/cmK$^2$) at room temperature, which is due to the complicated phonon dispersion arising from the misfit structure~\cite{currat}.

In contrast to Bi$_2$Sr$_2$CaCu$_2$O$_{8+\delta}$, the excess oxygen in Bi-based Co oxide is not changed by annealing, indicating the difficulty of changing the carrier density in the CoO$_2$ layer. The formal valences of Co ion in the layered Co oxides are around 3.3$\sim$3.5+, and they do not change very much against partial cation substitutions. It is also difficult to design n-type layered Co oxides.

\begin{table}[tb]
\begin{center}
\caption{
Chemical composition and lattice parameter of the misfit cobalt oxide (taken from various papers) 
}
\vspace{0.4cm}
\label{tbl1}
\begin{footnotesize}
\begin{tabular*}{12cm}{ccccccccc}
\hline
\\

\multicolumn{3}{l}{[Bi$_{0.868}$SrO$_2$]$_2$[CoO$_2$]$_{1.820}$ $^{\mbox{{\tiny Ref.}}}$~\cite{raveau}} & & \multicolumn{5}{c}{} \\
\multicolumn{3}{c}{Composition} & & \multicolumn{5}{c}{Lattice constants (\AA)} \\
Bi (A) & Sr (B) & Co (C) & & a$_{RS}$& b$_{RS}$ & b$_{H}$& c& $\beta$ ($deg.$) \\
1.736 & 2 & 1.82 & & 4.904 & 5.112 & 2.8 & 14.928 & 93.45 \\
\hline
\\

\multicolumn{3}{l}{[CaCo$_{0.5}$O$_{1.5-x}$]$_2$[CoO$_2$]$_{1.62}$ $^{\mbox{{\tiny Ref.}}}$~\cite{masset}} & & \multicolumn{5}{c}{} \\
\multicolumn{3}{c}{Composition} & & \multicolumn{5}{c}{Lattice constants (\AA)} \\
Co (A) & Ca (B) & Co (C) & & a$_{RS}$ & b$_{RS}$ & b$_{H}$ & c & $\beta$ ($deg.$) \\
1 & 2 & 1.62 & & 4.837 & 4.556 & 2.818 & 10.833 & 98.06 \\
\hline
\\

\multicolumn{3}{l}{[Ca$_{2}$CoO$_{3}$]$_{0.62}$[CoO$_2$] $^{\mbox{{\tiny Ref.}}}$~\cite{miya2}} & & \multicolumn{5}{c}{} \\
\multicolumn{3}{c}{Composition} & & \multicolumn{5}{c}{Lattice constants (\AA)} \\
Co (A) & Ca (B) & Co (C) & & a$_{RS}$ & b$_{RS}$ & b$_{H}$ & c & $\beta$ ($deg.$) \\
1 & 2 & 1.62 & & 4.837 & 4.556 & 2.818 & 10.833 & 98.06 \\
\hline
\\

\multicolumn{3}{l}{[Ca$_{2}$CoO$_{3}$][CoO$_2$]$_{1.62}$ $^{\mbox{{\tiny Ref.}}}$~\cite{sebastien}} & & \multicolumn{5}{c}{} \\
\multicolumn{3}{c}{Composition} & & \multicolumn{5}{c}{Lattice constants (\AA)} \\
Co (A) & Ca (B) & Co (C) & & a$_{RS}$ & b$_{RS}$ & b$_{H}$ & c & $\beta$ ($deg.$) \\
1 & 2 & 1.62 & & 4.83 & 4.54 & 2.82 & 10.76 & 98.1 \\
\hline
\\

\multicolumn{3}{l}{Tl$_{0.4}$[Sr$_{0.9}$O]$_{1.12}$[CoO$_2$] $^{\mbox{{\tiny Ref.}}}$~\cite{TlCo}} & & \multicolumn{5}{c}{} \\
\multicolumn{3}{c}{Composition (EDS)} & & \multicolumn{5}{c}{Lattice constants (\AA)} \\
Tl (A) & Sr (B) & Co (A+C) & & a$_{RS}$ & b$_{RS}$ & b$_{H}$ & c & $\beta$ ($deg.$) \\
0.8 & 2 & 2 & & 4.94 & 5.01 & 2.80 & 11.38 & 97.8 \\
\hline
\\

\multicolumn{3}{l}{[Pb$_{0.7}$Sr$_{1.9}$Co$_{0.4}$O$_{3}$][CoO$_2$]$_{1.8}$ $^{\mbox{{\tiny Ref.}}}$~\cite{pelloquin}} & & \multicolumn{5}{c}{} \\
\multicolumn{3}{c}{Composition (EDS)} & & \multicolumn{5}{c}{Lattice constants (\AA)} \\
Pb (A) & Sr (B) & Co (A+C) & & a$_{RS}$ & b$_{RS}$ & b$_{H}$ & c & $\beta$ ($deg.$) \\
0.72 & 2 & 2.26 & & 4.938 & 5.023 & 2.802 & 11.525 & 97.81 \\
\hline
\\

\end{tabular*}

\begin{tabular*}{12cm}{cccccccccc}

\multicolumn{4}{l}{[Pb$_{0.7}$Hg$_{0.2}$Sr$_{1.9}$Co$_{0.2}$O$_{3}$][CoO$_2$]$_{1.8}$ $^{\mbox{{\tiny Ref.}}}$~\cite{pelloquin}} & & \multicolumn{5}{c}{} \\
\multicolumn{4}{c}{Composition (EDS)} & & \multicolumn{5}{c}{Lattice constants (\AA)} \\
Hg (A) & Pb (A) & Sr (B) & Co (A+C) & & a$_{RS}$ & b$_{RS}$ & b$_{H}$ & c & $\beta$ ($deg.$) \\
0.22 & 0.72 & 2 & 2.1 & & 4.943 & 5.028 & 2.805 & 11.617 & 97.78 \\
\hline
\\

\end{tabular*}

\vspace{-0.4cm}
\end{footnotesize}
\end{center}
\end{table}

It would be useful to point out a difference from the recent material design for new thermoelectric materials, in which a concept of ``rattling'' is applied to reduce the thermal conductivity~\cite{slack}. The essence of the concept is to introduce an atom called ``rattler'' which is weakly bound in an oversized atom cage. Such an atom undergoes large local inharmonic vibrations, somewhat independent of the other atoms in the crystal. The rattlers can lower the thermal conductivity, but simultaneously change the carrier density. Thus, another atom must be doped for compensation. In contrast, the layered Co oxide can be designed by changing the block layer without modifying the structure of the CoO$_2$ layer. We can reduce the thermal conductivity by choosing a proper block layer, leaving the electronic properties of the CoO$_2$ layer unchanged. 

We will point out some differences from the block layer of the high-$T_c$ Cu oxides. At present, combinations of different block layers are not realized in the layered Co oxide, and bilayer or trilayer Co oxides are not synthesized. Doping control by cation substitution is also difficult, and the parent insulator of the layered Co oxides is not synthesized either. Instead of them, the block layer can induce ``uniaxial'' chemical pressure to the CoO$_2$ layer through the lattice misfit, which will be discussed in the next section.

\section{Crystal Structure of the Bi-based Co Oxide}

As described above, it is found that the block-layer concept is applicable to the layered Co oxides. Hereafter, we will focus on the Bi-based Co oxides, and show how to improve the thermoelectric properties by tuning the block layer. In the Bi-based Co oxides, the block layer is strongly coupled with the CoO$_2$ layer via ionic bonding. Since these two layers show different rotational symmetries, the electronic states would be altered to induce in-plane anisotropy in physical properties. This is a unique feature for the layered Co oxide. For high-$T_c$ Cu oxides, there is no misfit structure, and in-plane anisotropy is much weaker. Moreover, the misfit structure would induce anisotropic chemical pressure, which enables us to investigate a pressure effect on the CuO$_2$ layer. 

The Bi$_2$Sr$_2$O$_4$ block layer can be substituted by various cations. Trivalent Bi can be partially substituted by divalent Pb up to 30 \%, and Sr can be fully replaced by Ba and Ca~\cite{tarascon}. The lattice constants of the block layer vary with these substitutions without changing the lattice parameters of the CoO$_2$ layer, which implies that the misfitness $m$=$b_{RS}$/b$_{H}$ can be controlled.

All the samples were grown by TSFZ method starting from the nominal compositions of Bi$_{2-x}$Pb$_x$Sr$_2$Co$_2$O$_y$ ($x$=0, 0.4, and 0.6) and Bi$_{2}$Ba$_2$Co$_2$O$_y$. Hereafter, we refer to the samples by the nominal composition of Pb ($x$=0, 0.4, and 0.6). The actual compositions of the samples are listed against the nominal composition in Table~\ref{tbl2}(a). They were analyzed through inductively coupled plasma-atomic emission spectroscopy (ICP) and energy dispersive X-ray analysis (EDX). The EDX and ICP data are in excellent agreement with each other. We found that the actual composition of Pb is roughly equal to the nominal composition. In all samples, the chemical composition of Co is smaller ($\approx$ 1.8) than that of Bi+Pb and Sr ($\approx$ 2.0), which is consistent with the data in Table~\ref{tbl1}. For Bi$_{2}$Ba$_2$Co$_2$O$_y$, the Bi content ($\approx$ 2.2) is slightly larger than that of Sr and Co ($\approx$ 2.0).

\begin{table}[tb]
\begin{center}
\caption{
(a) Chemical composition estimated from energy dispersive X-ray (EDX) analysis and inductively coupled plasma-atomic emission spectroscopy (ICP). (b) Lattice parameter determined from four-circle X-ray diffractometer. $b_{RS}$ and $b_H$ are the $b$-axis lengths of the rock-salt and the hexagonal layers, respectively. $b_H$ is determined by the electron diffraction patterns.
}
\label{tbl2}
\begin{footnotesize}
\begin{tabular*}{6.5cm}{lccccc}
\multicolumn{3}{l}{(a) Composition}\\ 
\\ \hline 
Sample & & Bi & Pb & Sr & Co \\ 
\hline \hline
EDX &&&&&\\
$x$=0 &  & 2.01   &   0 & 2.0 & 1.81 \\ 
$x$=0.4 &  & 1.54 & 0.39 & 2.0 & 1.79  \\ 
$x$=0.6 & & 1.55 & 0.51 & 2.0 & 1.72 \\ 
Bi$_{2}$Ba$_2$Co$_2$O$_y$& & 2.2 & 0 & 2.0 & 2.04 \\ 
\hline 
ICP &&&&&\\
$x$=0  & & 2.0 & 0 & 2.0 & 1.9 \\ %
$x$=0.4  &  & 1.67 & 0.38 & 2.0 & 1.83 \\ 
$x$=0.6  &  & 1.52 &  0.57 & 2.0 & 1.9  \\ 
\hline
\end{tabular*}
\vspace{0.4cm}

\begin{tabular*}{10cm}{lccccc}
\multicolumn{3}{l}{(b) Lattice parameter}\\ 
\\ \hline 
Sample & $a_{RS}$ (\AA) & $b_{RS}$ (\AA) &$c$ (\AA) & $b_H$ (\AA) & $\beta$ (deg.) \\ \hline 
\hline
$x$=0 &  4.937 & 5.405 & 29.875 & 2.8 & 93.554\\
$x$=0.4&   4.914  &  5.221&29.974&2.8&92.315  \\ 
$x$=0.6&  4.904&5.206&30.041&-&92.657   \\ 
Bi$_{2}$Ba$_2$Co$_2$O$_y$& 4.9 &5.6 & 30.815& 2.8 & -\\
\hline
\end{tabular*}
\vspace{-0.4cm}
\end{footnotesize}
\end{center}
\end{table}

To determine the lattice constants of the rock-salt Bi$_2$Sr$_2$O$_4$ layer, we performed four-circle X-ray diffraction (XRD) measurements (Cu K$_{\alpha}$ X-ray source). Since Co atoms have a large mass absorption coefficient for Cu $K_\alpha$ radiation, we cannot observe the hexagonal CoO$_2$ layer. Instead, $b_H$ was estimated from transmission electron microscope (TEM) diffraction patterns.

Table~\ref{tbl2}(b) shows the lattice constants $a_{RS}$, $b_{RS}$, $c$, and $b_H$ and the angle $\beta$ for Bi$_{2-x}$Pb$_x$Sr$_2$Co$_2$O$_y$ and Bi$_{2}$Ba$_2$Co$_2$O$_y$. With increasing Pb concentration, the $c$-axis length increases continuously from 29.84 \AA~ of $x$=0 to 30.04 \AA~ of $x$=0.6, owing to the larger ion radius of Pb$^{2+}$. $b_{RS}$ discontinuously shrinks from 5.38 \AA~ of $x$=0 to 5.22 \AA~ of $x$=0.4, whereas $a_{RS}$ is nearly unchanged (4.9 \AA). These results are in good agreement with a powder XRD measurement by Yamamoto $et$ $al.$~\cite{yamamoto}.

\begin{figure}
\centering{
\includegraphics[width=6cm]{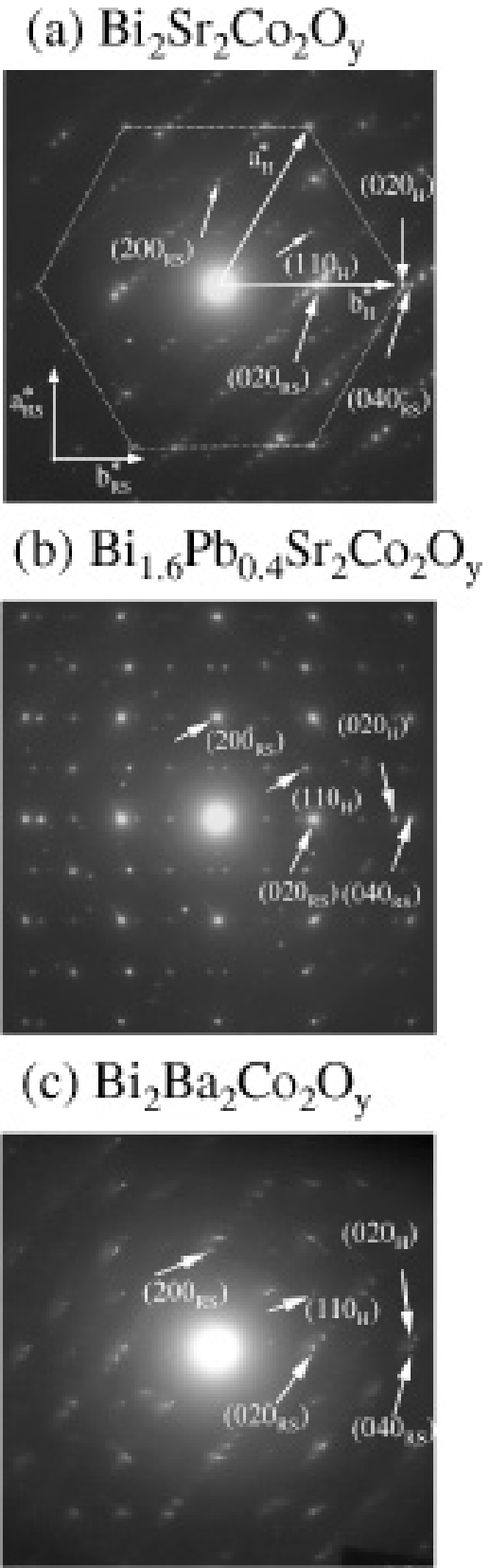}
\vspace{1cm}
}
\caption{
Electron diffraction patterns of (a) Bi$_{2-x}$Pb$_x$Sr$_2$Co$_2$O$_{y}$, $x$=0, (b) $x$=0.4, and (c) Bi$_2$Ba$_2$Co$_2$O$_y$.
 }
\label{fig3}
\end{figure}

TEM diffraction patterns of Bi$_{2-x}$Pb$_x$Sr$_2$Co$_2$O$_y$ ($x$=0, 0.4) and Bi$_2$Ba$_2$Co$_2$O$_y$ are shown in Figs.~\ref{fig3}(a)-(c), where the incident beam is parallel to the $c$ axis. The rock-salt diffraction pattern and the hexagonal diffraction pattern are clearly observed. $b_H$ is estimated to be about 2.8 \AA~ and the angle between them is about 60$^{\circ}$, which is independent of $x$ within the resolution limit of the TEM image. Note that the diffraction spots (040$_{RS}$) and (020$_{H}$) does not coincide each other due to the misfit structure.

In Bi$_{2-x}$Pb$_x$Sr$_2$Co$_2$O$_y$, 13 periods of $b_{H}$ roughly match with 7 periods of $b_{RS}$ (13 $b_H$ $\sim$ 7 $b_{RS}$ $\sim$ 36 \AA) while $a_{RS}$ $\sim$ $\sqrt{3}$ $a_{H}$ $\sim$ 4.9 \AA. The cation ratio of Bi : Sr : Co can be quantitatively understood by using the lattice parameters. The number of Bi or Sr atoms in 7 periods of the rock-salt block layer is 28, while 26 atoms of Co are in 13 periods of the CoO$_2$ layer. This indicates the cation ratio of Bi : Sr: Co = 2 : 2 : 1.85, which is consistent with the actual composition. As seen in the TEM diffraction pattern for Bi$_2$Ba$_2$Co$_2$O$_y$ [Fig.~\ref{fig3}(c)], the diffraction spots of (040$_{RS}$) and (020$_H$) appear at the same position, which indicates that the misfit structure disappears in this compound. The lattice constants are estimated to be $a_{RS}$=4.9 \AA~ and $b_{RS}$=5.6 \AA~, while $b_H$=2.8 \AA. The $c$-axis length is larger than that of Bi$_{2-x}$Pb$_x$Sr$_2$Co$_2$O$_y$ owing to the larger Ba ions. Since there is no incommensurate lattice constants between the rock-salt layer and the hexagonal layer, actual composition should be Bi$_2$Sr$_2$Co$_2$O$_y$, which contradicts with the EDX result. To clear this contradiction, precise structural analysis should be employed. 

\begin{figure}
\centering{
\includegraphics[width=12cm]{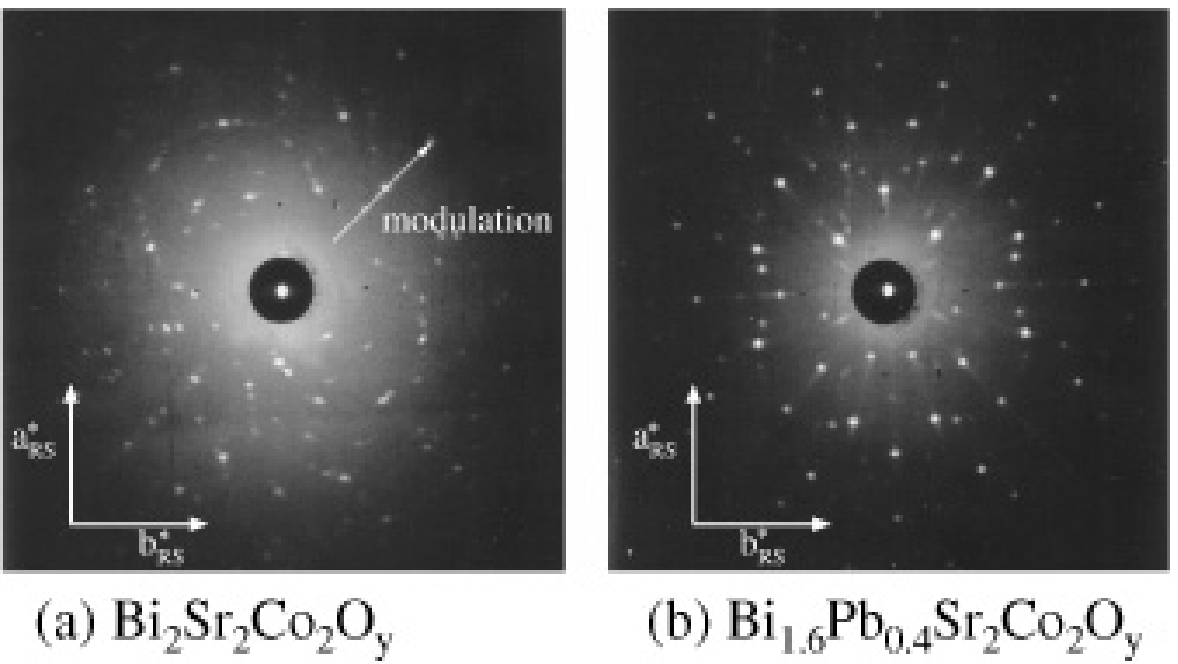}
\vspace{1cm}
}
\caption{ 
Laue transmission patterns of Bi$_{2-x}$Pb$_x$Sr$_2$Co$_2$O$_{y}$. (a) $x$=0 and (b) $x$=0.4.
 }
\label{fig4}
\end{figure}

The structural character in the Bi-based Co oxides is summarized as follows.

\begin{itemize}
\item[I. ] In Bi$_{2-x}$Pb$_x$Sr$_2$Co$_2$O$_{y}$, the lattice constants of the rock-salt block layer decrease discontinuously from $x$=0 to 0.4. This indicates the enhancement of the misfitness, because the lattice constant of the hexagonal CoO$_2$ layer does not change with $x$.
\item[II. ] In Bi$_{2}$Ba$_2$Co$_2$O$_{y}$, there is no trace of the misfit structure in the TEM diffraction pattern although it consists of the square block layer and the hexagonal CoO$_2$ layer.
\end{itemize}
Thus, we can control the misfit structure by substituting cations in the block layer.

Another feature in the Bi-based Co oxide is the highly modulated BiO layer, as is seen in high-$T_c$ Cu oxide Bi$_2$Sr$_2$CaCu$_2$O$_{8+\delta}$. The modulation runs along the $b$ axis~\cite{matsui1}, which disappears with Pb doping~\cite{matsui2}. The TEM diffraction patterns of $x$=0 and Bi$_{2}$Ba$_2$Co$_2$O$_{y}$ show satellite reflection due to the modulation structure along the oblique direction (tilted about 45$^\circ$) from $a^*_{RS}$ or $b^*_{RS}$. In contrast, there is no satellite reflection in $x$=0.4, indicating that the modulation structure disappears in this compound. This modulation structure was also confirmed with the Laue transmission photographs. Figs.~\ref{fig4}(a) and (b) show the Laue patterns of the $x$=0 and 0.4 samples, respectively. In Fig.~\ref{fig4}(a), a two-fold symmetry Laue pattern is observed, which is tilted about 45$^\circ$ from $a^*_{RS}$ or $b^*_{RS}$ axis. This two-fold symmetry Laue pattern is very similar to that of Bi$_2$Sr$_2$CaCu$_2$O$_{8+\delta}$~\cite{BSCCO}. On the other hand, the Laue pattern of the $x$=0.4 sample shows a four-fold symmetry, indicating that there is no modulation structure. 

Thus, the Bi-based Co oxides are very complicated with the misfit and modulation structures. We explain how the thermopower and the resistivity are affected by these structures in the next section.

\section{Effect of the Crystal Structure on the Thermoelectric Properties}

Before going into details of the transport properties, we will explain the electronic states of the edge-shared CoO$_6$ octahedra. Co $d$ orbitals with five-fold degeneracy split into the two-fold ($e_g$) degenerate and three-fold ($t_{2g}$) degenerate orbitals in the octahedral crystal field. In the edge-shared triangular CoO$_2$, the octahedra is compressed along the $c$-axis direction, and the $t_{2g}$ orbitals further split into the non-degenerate $a_{1g}$ and doubly degenerate $e'_g$ orbitals in the rhombohedral crystal field. The $e'_g$ orbitals spread along the CoO$_2$ layer to make a relatively broad band, while the $a_{1g}$ orbital spreads along the $c$-axis direction to make a narrow band. The formal valence of Co in the layered cobalt oxides is between 3+ and 4+, which corresponds to the configuration of (3$d$)$^{5}$ and (3$d$)$^{6}$. Thus the highest occupied state is the $a_{1g}$ orbital. The narrow $a_{1g}$ band serves large density of states at the Fermi level, and it is responsible for the large thermopower. On the other hand, the band calculation of NaCo$_2$O$_4$~\cite{shin} shows that the broad $e'_g + a_{1g}$ band slightly touches the Fermi level, which provides small numbers of holes with high mobility. Of course, the electronic states of NaCo$_2$O$_4$ are not exactly the same as those of the Bi-based Co oxides, but can be a good approximation for them in the lowest order.

\begin{figure}
\centering{
\includegraphics[width=8cm]{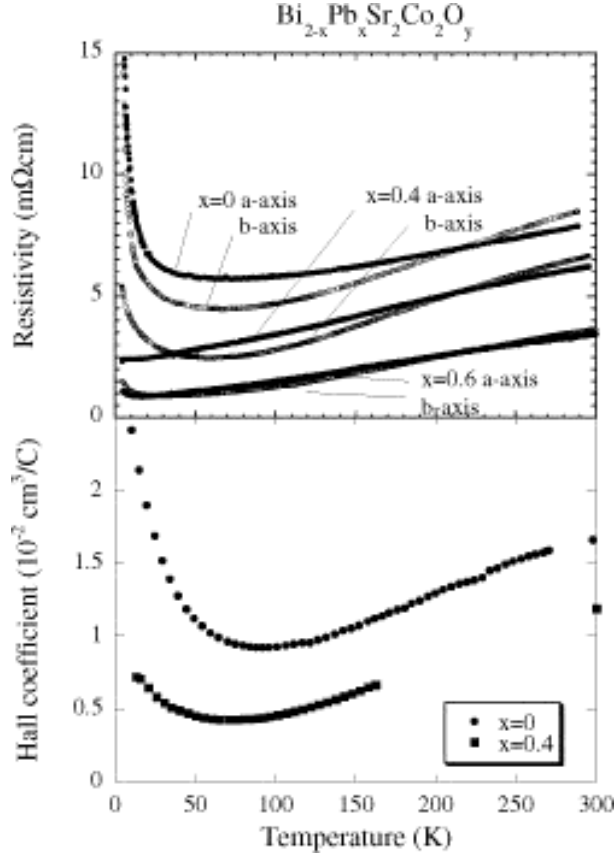}
\vspace{1cm}
}
\caption{ 
Temperature dependence of the (a) $a$- and $b$-axis resistivities and (b) Hall coefficient for Bi$_{2-x}$Pb$_x$Sr$_2$Co$_2$O$_y$.
}
\label{fig5}
\end{figure}

Figure~\ref{fig5}(a) shows the temperature dependence of the $a$- and $b$-axis resistivities for Bi$_{2-x}$Pb$_x$Sr$_2$Co$_2$O$_y$ (We refer the $a$- and $b$-axis directions as the $a_{RS}$ and $b_{RS}$ directions). All the samples show a metallic behavior at room temperature, implying an existence of the mobile carrier. The resistivity of $x$=0 shows minimum around 80 K and drastically increases with decreasing temperature. With increasing Pb concentration, the magnitude of the resistivity continuously decreases and the divergent behavior at low temperatures is strongly suppressed, indicating the carrier doping~\cite{yamamoto,ito}. The substitution of divalent Pb for trivalent Bi introduces 0.2 hole per Co site when Pb concentration increases from $x$=0 to 0.4. However, the decrease of resistivity from $x$=0 to 0.4 is smaller than that predicted from the chemical composition.

\begin{figure}
\centering{
\includegraphics[width=8cm]{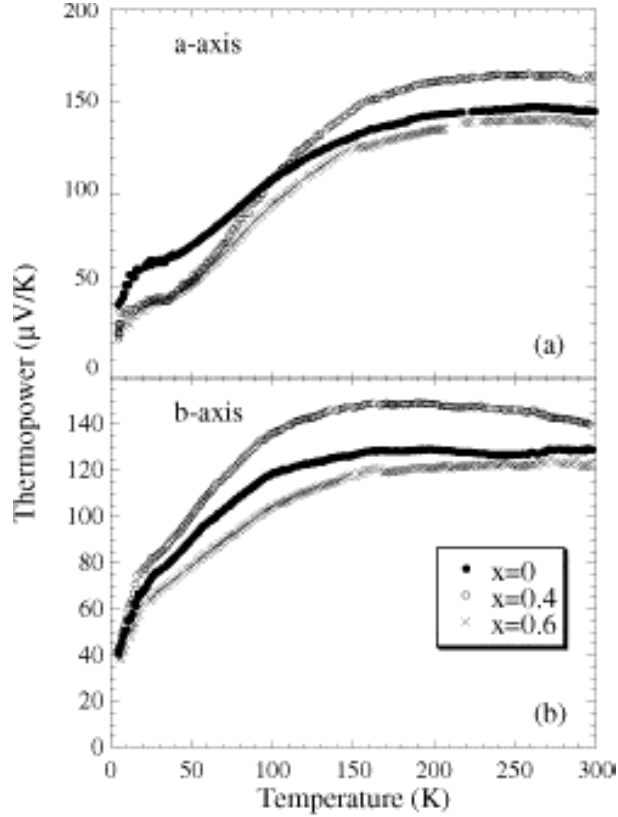}
}
\caption{
(a) $a$-axis thermopower and (b) $b$-axis thermopower for Bi$_{2-x}$Pb$_x$Sr$_2$Co$_2$O$_y$.
 }
\label{fig6}
\end{figure}

Figure~\ref{fig5}(b) shows the temperature dependence of the Hall coefficient ($R_H$) for $x$=0 and 0.4. Here, the current was applied parallel to the $a$ axis, and the magnetic field was applied perpendicular to the $ab$ plane. Since the simple relation $R_H$=1/$en$ gives a crude, but a reasonable estimation of carrier density, the decrease of $R_H$ with Pb doping from $x$=0 to 0.4 indicates the increase of the carrier density. The carrier densities at room temperature are 4$\times$10$^{20}$ cm$^{-3}$ (0.05 hole per Co site) for $x$=0 and 5$\times$10$^{20}$ cm$^{-3}$ (0.07 hole per Co site) for $x$=0.4. This indicates that only 0.02 holes per Co site are introduced with increasing Pb concentration from $x$=0 to 0.4, which is 1/10 of the value expected from the chemical composition. These results suggest that the carrier density does not change very much by cation substitution. Due to the existence of the $a_{1g}$ and $e'_g + a_{1g}$ bands, carrier-doping effects are complicated. Pb substitution not only dopes the carrier, but also modifies the electronic structure of the $a_{1g}$ band through the increase in the misfitness. One may notice that the temperature dependence of $R_H$ is very similar to that of the resistivity. The increase of $R_H$ and the resistivity at low temperature indicate the decrease of carrier density possibly due to the pseudogap formation, which we have previously proposed~\cite{ito}. 

Figures~\ref{fig6}(a) and (b) show the $a$- and $b$-axis thermopowers of various Pb concentrations, respectively. In both directions, the thermopower at room temperature increases with increasing Pb concentration from $x$=0 to 0.4, while, it decreases with further Pb substitution. This increase of the thermopower from $x$=0 to 0.4 cannot be explained by the conventional thermoelectric theory, where the thermopower decreases with increasing carrier density. Since the $b$-axis length discontinuously shrinks with Pb substitution from $x$=0 to 0.4, the increase of the thermopower is considered to arise from the chemical pressure. This behavior is very similar to the pressure dependence of the thermopower in the Ce-based compound~\cite{terra2, jaccard}. On the other hand, the decrease of the thermopower from $x$=0.4 to 0.6 is thought to be due to the increase of the carrier density, because the $b$-axis length is nearly unchanged.

\begin{figure}
\centering{
\includegraphics[width=9cm]{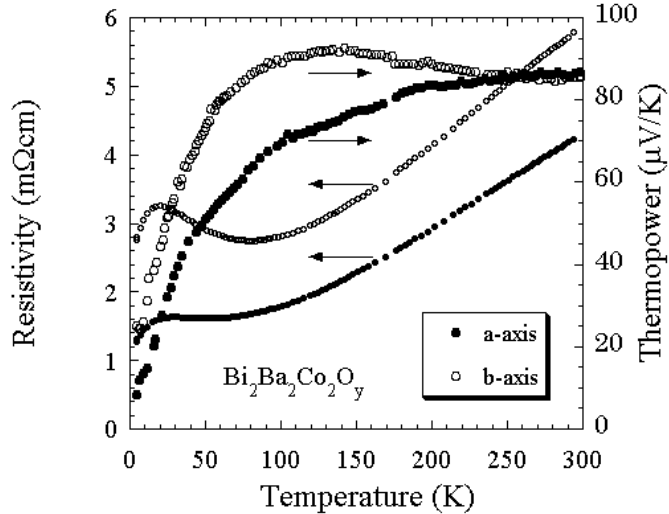}
}
\caption{
$a$- and $b$-axis resistivities and thermopowers for Bi$_{2}$Ba$_2$Co$_2$O$_y$.
 }
\label{fig7}
\end{figure}

Figure~\ref{fig7} shows the temperature dependence of the resistivity and thermopower for Bi$_2$Ba$_2$Co$_2$O$_y$. The resistivity of Bi$_2$Ba$_2$Co$_2$O$_y$ is more metallic than that of Bi$_2$Sr$_2$Co$_2$O$_y$~\cite{tarascon,ito2}, and its magnitude is as small as that of the $x$=0.6 ($\sim$ 3 m$\Omega$cm at 300 K). The thermopower at room temperature is much smaller than that of Bi$_2$Sr$_2$Co$_2$O$_y$, and is about 90 $\mu$V/K. These results indicate that Ba and Pb substitutions cause the hole doping, but the effects on the thermopower are different. The decrease of the thermopower can not be explained solely by the increase of the carrier density, because the thermopower of Bi$_2$Ba$_2$Co$_2$O$_y$ is about 60 $\mu$V/K smaller than that of $x$=0.6. Since the misfit structure is not seen in the TEM diffraction pattern of Bi$_2$Ba$_2$Co$_2$O$_y$, the decrease of the thermopower is thought to be due to the disappearance of the misfit structure. 

\begin{figure}
\centering{
\includegraphics[width=8cm]{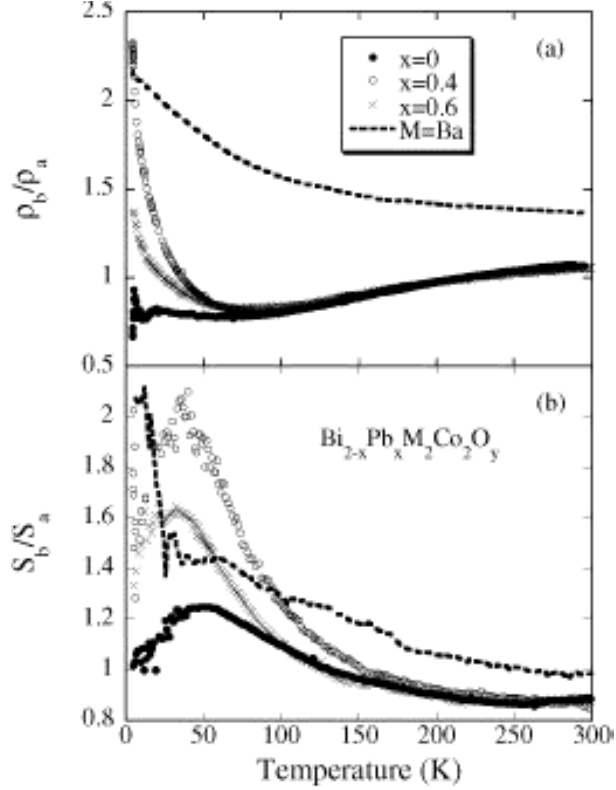}
\vspace{1cm}
}
\caption{
In-plane anisotropy of the (a) resistivity ($\rho_b$/$\rho_a$) and (b) thermopower ($S_b$/$S_a$) for Bi$_{2-x}$Pb$_x$Sr$_2$Co$_2$O$_y$ and Bi$_{2}$Ba$_2$Co$_2$O$_y$. 
 }
\label{fig8}
\end{figure}

Next, we will discuss the in-plane anisotropy in the resistivity ($\rho_b$/$\rho_a$) and the thermopower ($S_b$/$S_a$). $\rho_b$/$\rho_a$ and $S_b$/$S_a$ of the various Pb contents are shown in Figs.~\ref{fig8} (a) and (b), respectively. The in-plane anisotropy of Bi$_2$Ba$_2$Co$_2$O$_y$ is also shown in the same figure by a dotted line. In all doping levels, the resistivities and the thermopowers show an in-plane anisotropy, which does not expect in the hexagonal symmetry CoO$_2$ layer. In spite of the absence of misfit structure, $\rho_b$/$\rho_a$ and $S_b$/$S_a$ for Bi$_2$Ba$_2$Co$_2$O$_y$ show large in-plane anisotropy. Note that $S_b/S_a$ is roughly equal to that of $\rho_b/\rho_a$. These results strongly suggest that the in-plane anisotropy in the resistivity and the thermopower have the same origin. We attribute the large in-plane anisotropy to the anisotropic pseudogap formation. Due to the symmetry difference between the rock-salt layer and the hexagonal layer, the crystal symmetry of the CoO$_2$ layer becomes lower to stabilize the spin-density-wave-like state as was discussed previously~\cite{fujii1}. Thus, the anisotropic pseudogap would be open and reduces the density of states. Another possibility to explain the large in-plane anisotropy is the self-organization of the electronic system such as stripe order seen in high-$T_c$ Cu oxides.

\begin{figure}
\centering{
\includegraphics[width=9cm]{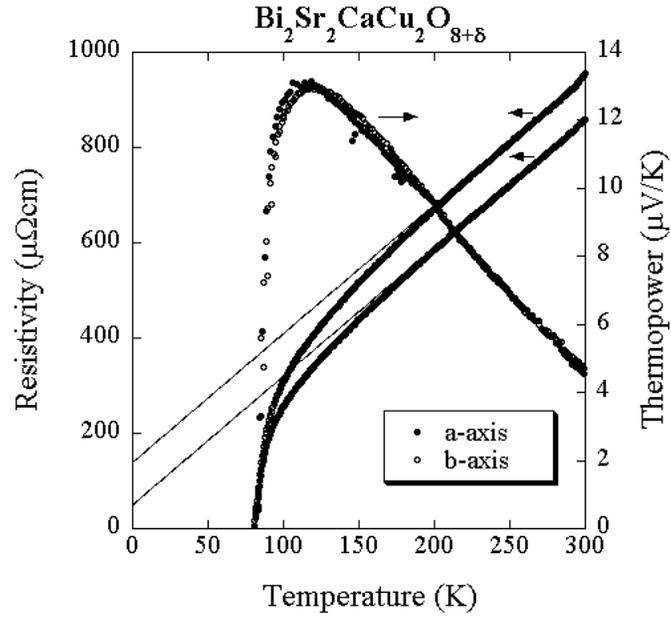}
}
\caption{
Temperature dependence of the (a) resistivity ($\rho_b$, $\rho_a$) and (b) thermopower ($S_b$, $S_a$) for Bi$_{2}$Sr$_2$CaCu$_2$O$_{8+\delta}$. 
 }
\label{fig9}
\end{figure}

Contrary to this system, the high-$T_c$ Cu oxide Bi$_2$Sr$_2$CaCu$_2$O$_{8+\delta}$ has the rock-salt Bi$_2$Sr$_2$O$_4$ layer and the square CoO$_2$ layer. Figure~\ref{fig9} shows the resistivities and thermopowers of Bi$_2$Sr$_2$CaCu$_2$O$_{8+\delta}$ along the $a$ and $b$ axes. The resistivities show rather small in-plane anisotropy compared to Bi$_2$Sr$_2$Co$_2$O$_y$. $\rho_b$ is higher than $\rho_a$, where $\rho_{b}$ is roughly expressed as a sum of $\rho_{a}$ and a $T$-independent residual resistivity ($\rho_{0}$) at high temperature. Typical negative slope of the thermopower for high-$T_c$ Cu oxide was observed. In contrast to the in-plane resistivities, there is no anisotropy in the thermopower within our experimental error. These results suggest that the modulation structure along the $b$ axis works as an anisotropic scattering center~\cite{fujii}, because the thermopower does not depend on the scattering rate in the lowest order approximation. Thus, with the stack of the same symmetry layers, we cannot observe significant in-plane anisotropy both in the resistivity and in the thermopower.

\begin{figure}
\centering{
\includegraphics[width=8cm]{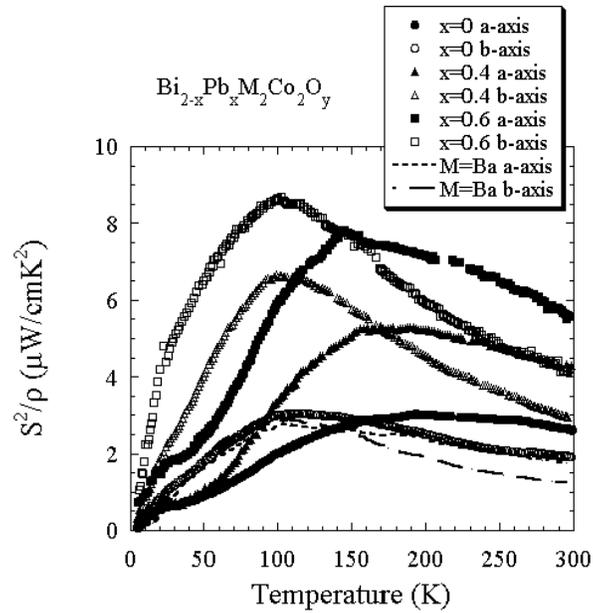}
}
\caption{
The power factor ($S^2/\rho$) of Bi$_{2-x}$Pb$_x$Sr$_2$Co$_2$O$_y$ and Bi$_{2}$Ba$_2$Co$_2$O$_y$ along the $a$- and $b$- directions.
 }
\label{fig10}
\end{figure}

One may notice that for Bi$_2$Sr$_2$Co$_2$O$_y$ and Bi$_2$Ba$_2$Co$_2$O$_y$, the anisotropy of the resistivity is weaklier dependent on temperature than that of the thermopower. We attribute this to the modulation structure, and think that the anisotropy is averaged by the modulation structure, whose direction is tilted by 45$^\circ$ from the $a$- and $b$-axes.

We briefly mention the thermoelectric performance in the Bi-based Co oxide. The power factor ($S^2/\rho$), which is a measure of the thermoelectric properties, of Bi$_{2-x}$Pb$_x$M$_2$Co$_2$O$_y$ is shown in Fig.~\ref{fig10}. The magnitude of the power factor increases with increasing Pb content, reflecting the reduction of the resistivity. The power factor along the $b$ axis is more than two times as large as that along the $a$ axis near 50 K. (Fig.~\ref{fig10}), while the power factor of Bi$_2$Ba$_2$Co$_2$O$_y$ shows no anisotropy at low temperature. This indicates that the thermoelectric properties can be improved (doubled in the present case) with controlling the misfit structure. The thermopower and resistivity for $x$=0.4 are about 150 $\mu$V/K and 5 m$\Omega$cm at 300 K, which is the same order to those of Bi$_2$Te$_3$ (200 $\mu$V/K and 200 $\mu$$\Omega$cm at 300 K). The optimum power factor is as large as 9 $\mu$W/cmK$^2$ around 50 K for $a$-axis of $x$=0.6, which is highest for thermoelectric oxides at 100 K. In Bi$_2$Ba$_2$Co$_2$O$_y$, the power factor was lower than that of Bi$_2$Sr$_2$Co$_2$O$_y$, where the decrease in thermopower is more serious than the reduction of the resistivity. 

\section{summary}

In this article, we have proposed the block-layer concept for a material design for thermoelectric oxides, where the crystal symmetry, the chemical pressure, the carrier concentration are controlled without damaging the structure of the CoO$_2$ layer. This makes a remarkable contrast to the phonon-glass concept, the state-of-the-art design for thermoelectric semiconductor, in which rattlers inevitably affect the conductive framework of the host. We can also say that the block-layer concept proposed here is a concept to build a new function by combining different blocks responsible for different functions, just like a functional group in organic chemistry and/or a protein design in bio-engineering. In this sense, thermoelectrics can be a fertile playground, in the sense that the three transport parameters are to be harmonized to achieve a large value of $ZT$. Currently new layered Co oxides (and equivalently new block layers) have been synthesized one after another, which enables us to design a thermoelectric oxide with high flexibility. We hope that the best combination of the block layers will be discovered in near future, and also hope that oxide thermoelectrics will be realized by using the layered Co oxides.

\subsection*{Acknowledgments}

The authors would like to thank A. Matsuda and T. Watanabe for their collaboration at the early stage of this work, and also appreciate T. Goto and S. Enomoto for crystal characterization.

\end{document}